\documentclass[cits]{PoS}

\usepackage{graphicx}

\def\phcmskeV{\mbox{ph~cm$^{-2}$\,s$^{-1}$\,keV$^{-1}$}}

\newcommand{\umrm}[1]{^{\mathrm{#1}}}

\title{Analysis of the new INTEGRAL Earth observations to measure the cosmic X-ray background}

\ShortTitle{Analysis of the new INTEGRAL Earth observations to measure the cosmic X-ray background}

\author{\speaker{M. T\"urler}, N. Produit, L. Pavan, C. Ferrigno, and P. Bordas\thanks{also at: Institut f\"ur Astronomie und Astrophysik, Universit\"at T\"ubingen, Sand 1, 72076 T\"ubingen, Germany}\\
        ISDC \& Observatory of the University of Geneva, ch. d'Ecogia 16, 1290 Versoix, Switzerland\\
        E-mail: \email{marc.turler@unige.ch}
}

\abstract{%
A new series of Earth occultation observations have been started in 2012 to
refine the determination of the cosmic X-ray background by the \emph{INTEGRAL}
mission. We show here that the new detector lightcurves in the
3 to 160 keV range differ from the ones obtained in 2006.
Instead of the expected modulation induced by the passage of the Earth through
the field of view of the JEM-X, IBIS/ISGRI and SPI instruments, we record unrelated
variability on shorter timescales.

We discuss the differences obtained with the datasets of 2006 and 2012 in view of
the changes in pointing direction, spacecraft orbit and solar cycle phase.
We conclude that the Earth occultation signal in 2012 is likely blended by
radioactive decay resulting from the activation of the spacecraft when crossing the
proton radiation belt at perigee passage. The observed variability, on the other hand,
results most likely from the current solar maximum. In addition to a variable particle environment
from inhomogeneities of the solar wind, we also find evidence for hard X-ray auroral
emission. While the former can be traced by SPI/ACS counts, the latter -- by enhancing unpredictably
the Earth emission -- is a major disturbance for measuring the diffuse X-ray background
through occultation by the Earth.
}

\FullConference{"An INTEGRAL view of the high-energy sky (the first 10 years)"
9th INTEGRAL Workshop and celebration of the 10th anniversary of the launch,\\
		October 15-19, 2012\\
		Biblioth\`eque Nationale de France, Paris, France}

\begin{document}

\section{Introduction}
\label{sec:intro}
The idea to use the Earth disk as a shield to occult the cosmic X-ray background
(CXB) was evaluated early in the \emph{INTEGRAL} mission and was implemented in January
and February 2006. At the start of four spacecraft revolutions, the Earth was allowed
to cross the field of view (FoV) of the instruments aimed at a fixed position in the sky.
The occultation of the CXB by the Earth passage resulted in a characteristic
signal, a hollow, in the evolution of the detector count rates.
The depth of the signal modulation in different energy bands can then be used to extract the
spectrum of the CXB. Based on the successful performance and analysis of these very
unconventional observations \cite{CSR07,TCC10} , it has been proposed in
2010 to repeat the experiment.

The new set of observations was aimed at deriving the spectrum of the CXB in the hard
X-ray domain with improved statistics and less systematics. The idea was to gain
a factor of two in statical uncertainties by multiplying by four the number of occultations.
The gain in systematics was expected by performing them in a FoV not affected by the
presence of the Galactic ridge and bright point sources as it was the case in 2006.
As the bombardment of the Earth by cosmic rays is reduced during solar maximum, it was
also anticipated to have a lower emission by the Earth atmosphere above 70 keV in 2012--2014
with respect to the solar minimum of 2006 \cite{SCS07}. This strong scientific case and the
evaluation of the feasibility of the observations by the staff of the Mission Operation Centre \cite{HLS11}
led to the recommendation by the \emph{INTEGRAL} Users Group to perform the proposed
observation program.

Four new Earth occultation observations have been conducted from May 2012 to September 2012.
We present here these observations for all three high-energy instruments aboard \emph{INTEGRAL}.
The obtained data presented in Sect.~\ref{sec:data} do unfortunately not match the expectations
for various reasons that we discuss in Sect.~\ref{sec:discussion}.

\begin{figure}
\centerline{\includegraphics[width=.9\textwidth]{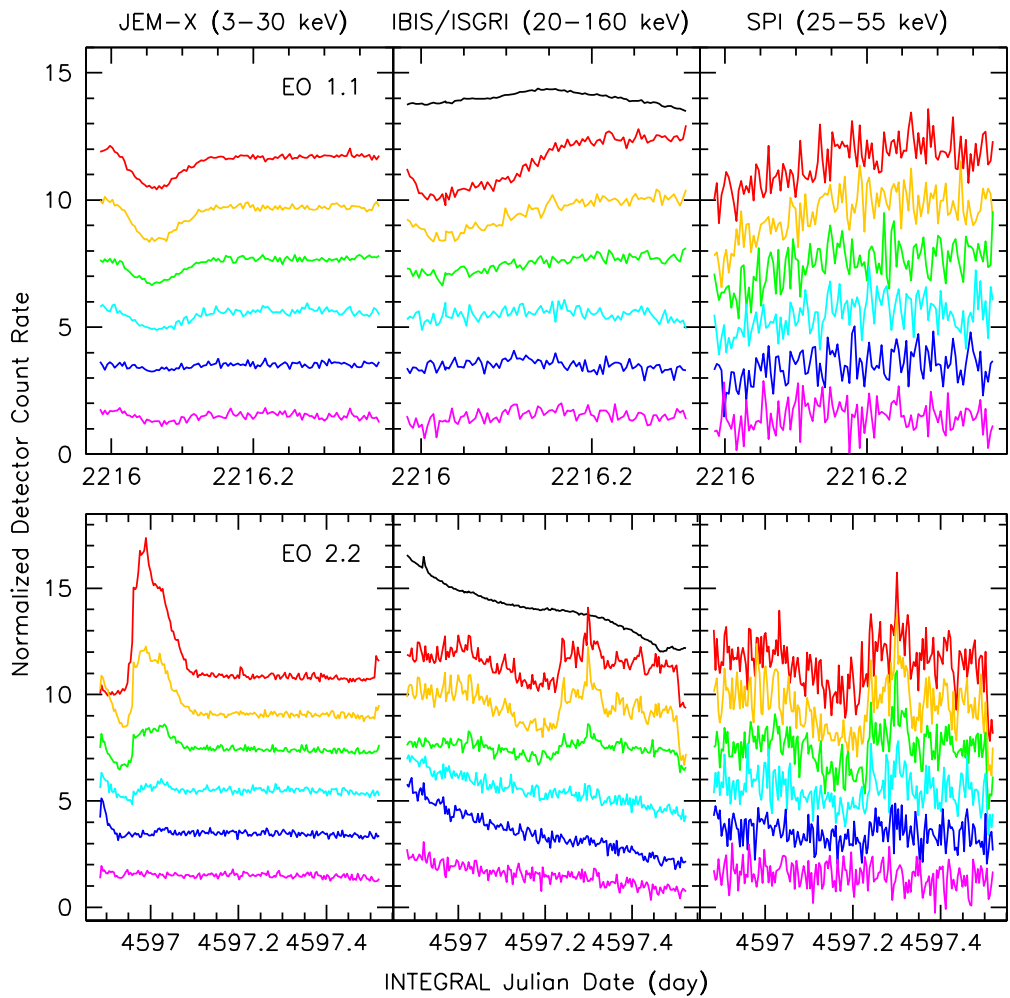}}
\caption{%
\label{fig:1}
   Comparison of JEM-X, IBIS/ISGRI and SPI detector lightcurves for Earth observations
   EO 1.1 (upper panel) and EO 2.2 (lower panel). For each instrument, we show six
   average-subtracted lightcurves in different energy bands with energies increasing
   from the top curve (red) to the bottom curve (magenta). The SPI lightcurves
   have also been divided by the square-root of the average count rate to better match
   JEM-X and IBIS rates. JEM-X lightcurves are the average of JEM-X 1 and 2 in the energy bands:
   3--5, 5--8, 8--12, 12--18, 18--24, 24--30 keV.  ISGRI and SPI bands are: 20--28, 28--38, 38--52,
   52--72, 72--100, 100-160 keV and 25--30, 30--35, 35--40, 40--45, 45--50, 50--55 keV, respectively.
   The simultaneous SPI/ACS lightcurve variability is also shown as a black line for comparison with
   IBIS counts.
}
\end{figure}

\section{Data}
\label{sec:data}
The first four new \emph{INTEGRAL} Earth-occultation observations (EOs) of 2012 have been conducted at the start
of the spacecraft revolutions 1169 (May 10), 1197 (Aug. 01), 1206 (Aug. 28) and 1216 (Sep. 27).
We refer to them as EO 2.1, 2.2, 2.3 and 2.4, respectively. We extracted full detector lightcurves
for all three high-energy instruments aboard \emph{INTEGRAL} \cite{WCD03}. We used version 10 of
the Off-line Scientific Analysis (OSA 10) provided by the ISDC \cite{CWB03} with a small change
for the JEM-X instrument to allow the analysis of slew observations rather than pointings.
For IBIS/ISGRI we analysed the data until the deadtime correction level and extracted the lightcurves
with the \texttt{ii\_light} executable included in the OSA package in time bins of 300\,s.
For SPI, we simply selected the events of all live detectors based on time and energy.

Figure~\ref{fig:1} shows one of the new sets of detector lightcurves (EO 2.2) compared
to those obtained the same way using the data of EO 1.1 in revolution 401 (Jan. 2006).
The drop in the count rates resulting from the passage of the Earth through the FoV of the instruments
is clearly visible in the lowest energy lightcurves of each instrument for EO 1.1, but is not seen for EO 2.2.
For the latter observation, a CXB occultation might be present in JEM-X at the very beginning, before a
strong emission bump arises in the signal. The SPI and IBIS lightcurves at energies below 60 keV
show correlated variability which is not matching the evolution of the SPI/ACS counts.

\begin{figure}
\centerline{\includegraphics[bb=16 144 590 610, clip, width=.9\textwidth]{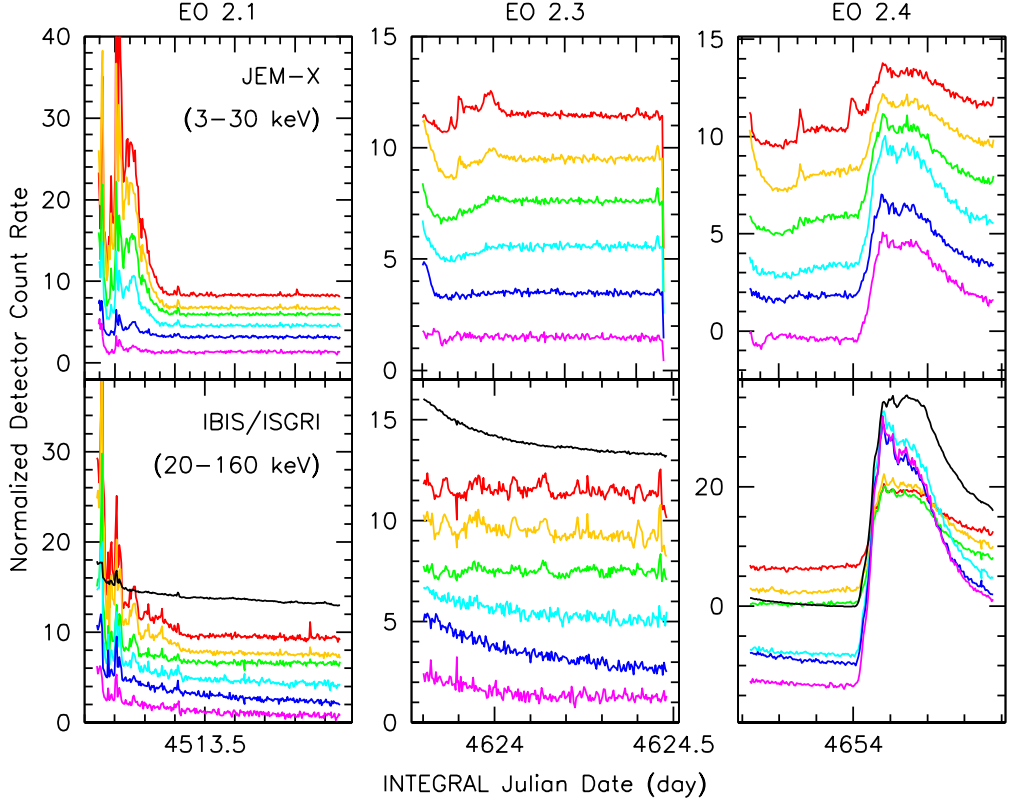}}
\caption{%
\label{fig:2}
   Comparison of JEM-X (upper panels) and IBIS/ISGRI (lower panels) detector lightcurves
   for Earth observations EO 2.1, EO 2.3 and EO 2.4. The energy bands and colour coding is
   as in Fig.~1.
}
\end{figure}

The detector lightcurves of the three other EOs of 2012 do also not display the expected
occultation modulation, as shown in Fig.~\ref{fig:2}. While the lightcurves of EO 2.3 are relatively stable,
those of EO 2.1 and EO 2.4 are affected by strong variability also visible in the SPI/ACS count rates.
 
\section{Discussion}
\label{sec:discussion}
The diversity of the detector lightcurves of the EOs of 2012 and the absence
of the expected CXB occultation signal present in the data of 2006 is puzzling. In order to understand
the problem, we discuss below the possible effect of three main differences between the EOs of 2012
and the ones of 2006: a change in pointing direction, in spacecraft orbit, and in the solar cycle phase.

\subsection{Effect of the pointing direction}
The EOs of 2006 were directed at a Galactic longitude $l\approx -30\umrm{o}$ not far from the
Galactic bulge and including the Galactic ridge X-ray emission (GRXE) in the partially coded FoV of IBIS.
This was not very favorable, as it resulted also in the occultation of the GRXE and of bright point
sources in addition to the CXB occultation. The new observations are still pointing close
to the Galactic plane, but at $l\approx -110\umrm{o}$ where the GRXE is much
reduced \cite{KTR12}.

This better configuration with respect to the background sky could however imply an increase
of the albedo of the Earth by atmospheric reflection of the Galactic bulge X-ray
emission. The time-averaged contribution of all point sources in the central radian of the Milky Way
($|l|\!<\!30\umrm{o}$ and $|b|\!<\!15\umrm{o}$) has been derived from SPI observations to be a
powerlaw above 20\,keV with a photon index $\Gamma$\,=\,2.67 and a flux at 50\,keV of
2.47\,$\times$\,10$^{-3}$ \phcmskeV\ \cite{BJR08}. As this normalization is just about the flux of
the Crab Nebula at 50\,keV -- which is located in the opposite direction on the sky -- the Earth albedo
due to point sources should be roughly constant for any pointing direction. Point sources contribute
 thus only an additional $\sim$2--3\,\% (1/40) to the albedo of the Earth from CXB reflexion, as
 estimated with the Crab flux for the EOs of 2006 \cite{CSR07}.

\subsection{Effect of the spacecraft orbit}
Another difference with respect to the EOs of 2006, is the change in the orbit of the spacecraft. The perigee height
during the first four EOs of 2012 was only between $\sim$3300\,km and $\sim$3800\,km, whereas it was
at 12'600\,km for the EOs of 2006. As a consequence, \emph{INTEGRAL} is currently passing through the proton
radiation belt at perigee, which leads to activation of the spacecraft material that is then radioactively
decaying during the post-perigee EOs. This leads to an exponential decay of the instrumental background that
is clearly present in the SPI/ACS detector count rates and in the highest energy bands of IBIS/ISGRI
(see Figs.\ref{fig:1} \& \ref{fig:2}).

In parallel to this change of perigee height, the orbit precessed such that in 2012 the exit of the electron
radiation belt is at about twice higher altitude, $\sim$50'000\,km, than the belt entrance height, whereas
in 2006 we had the opposite situation with an exit at $\sim$35'000\,km and an entrance at $\sim$60'000\,km.
The higher exit height in 2012 delays the start of the EOs after perigee and thus results in a slightly smaller Earth
size and a slower passage of the Earth through the FoV of the instruments. As a consequence, the EOs of 2012 do
only start when the Earth is already close to the centre of the IBIS FoV and their duration is about twice
longer ($\sim$16h) than in 2006 ($\sim$8\,h).

The overall consequence of this change of orbit for post-perigee EOs is that the expected Earth modulation
signal in IBIS and SPI is basically a continuous increase of the count rates during the observations.
This rise of detector counts can be easily blended by the opposite decrease
of counts resulting from the radioactive decay of  the spacecraft. We would need to analyse several
beginning of revolutions -- preferentially in clear extragalactic fields -- to evaluate if the contribution of the
radioactive decay of the spacecraft to the detector lightcurves in different energy bands can be uniquely
assessed and removed from the EOs. This would also need to be done for JEM-X which is known to have 
a variable gain in the first few science windows after perigee passage. The possibility of performing
pre-perigee observations would solve all these issues.

\subsection{Effect of the solar cycle}
A third difference with respect to the EOs of 2006, is that we are now close to solar maximum. This was
initially seen positively (see Sect.~\ref{sec:intro}), but turns out to result in complex variability in the
detector counts. Variations in the solar wind experienced by \emph{INTEGRAL} result in a variable
instrumental background that can be traced by the SPI/ACS detector and the IREM electron count rates.
This kind of variability does strongly affect EOs 2.1 and 2.4 (see Fig.~\ref{fig:2}).

The variability seen in EO 2.2 (Fig.~\ref{fig:1}) is likely of another kind. In the JEM-X bands it seems
quite clear that we observe an event of strong auroral activity \cite{BER07} with a soft spectrum creating
the bump of emission when the Earth polar regions are in the FoV of the instrument. The slight enhancement
seen quasi-simultaneously in the lowest energy bands of IBIS and SPI might be related. The other
variability event in the second part of the EO is only seen in IBIS and SPI, but
not in the SPI/ACS and in JEM-X, suggesting that it is not instrumental, but rather also auroral emission
that occurred when the Earth has already left the smaller FoV of JEM-X. We note that the recent EO 2.5
(not presented here) shows a similar behaviour with much higher amplitude. It might be of interest to
study hard X-ray auroral events from space, although their variability and spectral characteristics are
already well known from balloon experiments back in the 1960s \cite{BR66}. Concerning the extraction
of the CXB spectrum, aurorae are clearly a serious nuisance, especially if their variability
timescale is of the order of several hours, as is the occultation signal.

\section{Conclusion}
\label{sec:conclusion}

The detector lightcurves of the four first EOs of 2012 do not show the expected modulation
from the passage of the Earth through the FoV of the instruments, thus preventing us from
extracting the spectrum of the CXB. The occultation signal is likely blended by
radioactive spacecraft decay after radiation belt exit. This issue could be solved by performing
pre-perigee observations. Unfortunately, the data are also contaminated by variations
due to the current strong solar activity. This has both an effect on the instrumental
background due to a variable solar wind, and on the auroral X-ray emission
of the Earth. The latter effect might be interesting in itself, but is a major issue for
determining the CXB spectrum by Earth occultations at solar maximum.

\acknowledgments{
We are very grateful to the \emph{INTEGRAL} staff at ISOC and MOC for scheduling and performing these unconventional observations.
}

\end{document}